**Three-Dimensional Printed Liquid Diodes with Tunable Velocity: Design Guidelines and Applications for Liquid Collection and Transport**

*Camilla Sammartino,¶ Michael Rennick,¶ Halim Kusumaatmaja,\* Bat-El Pinchasik \*\**

C.S and B.-E.P.

Tel-Aviv University

School of Mechanical Engineering

Faculty of Engineering

6997801

Tel-Aviv, Israel

M.R and H.K

Department of Physics

Durham University South Road

Durham DH1 3LE

United Kingdom

¶ These authors contributed equally to the work.

**Corresponding Authors**

\* halim.kusumaatmaja@durham.ac.uk

\*\* pinchasik@tauex.tau.ac.il

**Abstract**

Directional and self-propelled flow in open channels has a variety of applications, including microfluidic and medical devices, industrial filtration processes, fog-harvesting and condensing apparatuses. Here, we present versatile three-dimensional (3D)-printed liquid diodes that enable spontaneous unidirectional flow over long distances for a wide range of liquid contact angles. Typically, we can achieve average flow velocities of several millimeters per second over a distance of tens to hundreds of millimeters. The diodes have two key design principles. First, a sudden widening in the channels' width, in combination with a small bump, the pitch, ensure pinning of the liquid in the backward direction. Second, an adjustable reservoir, the bulga, is introduced to manipulate the liquid velocity with differing expansion angles. Using a combination of experiments and lattice Boltzmann simulations, we provide a comprehensive analysis of the flow behavior and speed within the channels with varying contact angles (CA), pitch heights and bulga angles. This provides guidelines for the fabrication of bespoke liquid diodes with optimal design for their potential applications. As a feasibility investigation, we test our design for condensation of water from fog and subsequent transport uphill.

**1. Introduction**

Many industrial applications rely on directional transport of liquids, such as microfluidics,[1] printing,[2] filtering[3] and lubrication apparatuses,[4] water-harvesting,[2] medical devices[5] and precise irrigation systems.[6] While existing technologies feature external energy sources such as micropumps and moving parts to drive the flow, they also increase production and operating costs and add potential for malfunctions.[7] However, passive solutions to this problem, including fluid rectifiers such as Tesla valves,[8] require high Reynold's numbers or non-Newtonian fluids.[9]

In recent years, the fabrication of devices that are capable of spontaneous unidirectional liquid transport attracted much attention. Such devices, named liquid or fluidic diodes in analogy with the electronic counterpart,[10] exploit capillary action as the driving force. The key is to create a pressure difference that favors flow in one direction, while halting it in the opposite one. This can be done by introducing surface wettability gradients, leading to so-called slip-mode[11] liquid diodes. Alternatively, we can harness the geometry of the solid surface, in particular by employing asymmetric structures, to design spreading-mode[11] liquid diodes. For example, sudden openings that increase the liquid interfacial area can halt motion due to the increased energy required to overcome these features, while inducing motion in the opposite direction. Indeed, liquid motion in spreading-mode diodes often demonstrate stick-slip behavior[12] where an advancing liquid front is paused until these geometric features are overcome. It is also worth noting that several of the works conducted on liquid diodes are inspired by nature,[13] exhibiting passive unidirectional transport of liquids with remarkable efficiency and precision. These include fleas,[14] insects,[15] lizards,[16] butterflies,[17] spider silk,[18] cacti,[19] pitcher plants[20] and the beak of shorebirds.[21]

Despite the continuous growth of this field, further developments are needed to improve the performance of these passive devices in terms of flow rate and directionality, and provide general design guidelines that are adaptable to multiple use cases.[2] To this end, we present 3D printed spreading mode liquid diodes with a highly modifiable design. Importantly, we demonstrate for the first time how the two structural features, the pitch and the bulga, can be exploited to separately control the unidirectionality and velocity of the liquid flow. Our design allows long distance transport of tens to hundreds of millimeters, with a typical speed of several millimeters per second.

3D printing has been on the rise over recent years in several fields for large-scale production.[22] Here, the use of 3D printing allows us to easily adjust key structural features in the diode design. This economic and accessible choice may replace other frequently employed techniques such as laser cutting,[14] self-assembled monolayers [23,24] lithography,[25] etching[26] or additional delicate post-treatments.[27] These methods are more expensive and laborious, with potential for irregularities and durability problems in the long term.[28]

We use a combination of experiments and lattice Boltzmann simulations to provide a comprehensive, systematic analysis on the performance of the liquid diodes as we vary the aforementioned structural features and the surface tension (or the contact angle) of the liquid used. We demonstrate that the pitch, designed as an elliptic cylindrical bump, widens the working window of contact angles by introducing more degrees of freedom in the system, and that the bulga allows fine control over the velocity by controlling the volume of an additional reservoir. Finally, we test our design for condensation of water from fog and subsequent horizontal transport or uphill.

## 2. Results and Discussion

### 2.1. Design

Fig. 1 shows an overview of the diode design. Each diode typically consists of a unit cell, repeated 14 times, and 2 reservoirs, one at each end (Fig. 1a, i). Longer diodes, consisting of 30 unit cells with total length of 146.4 mm, were also printed and successfully tested (see ESI, Video S1). Each unit cell features a central triangular area, the bulga, and entrance and exit paths, the hilla and the orifice, respectively. The pitch is located at the exit of the bulga (Fig. 1a, ii). Lattice Boltzmann simulations allowed us to compare the performance of different pitch shapes. The elliptic cylinder provides the best compromise between diodicity and ease of fabrication (Fig. S1, ESI).

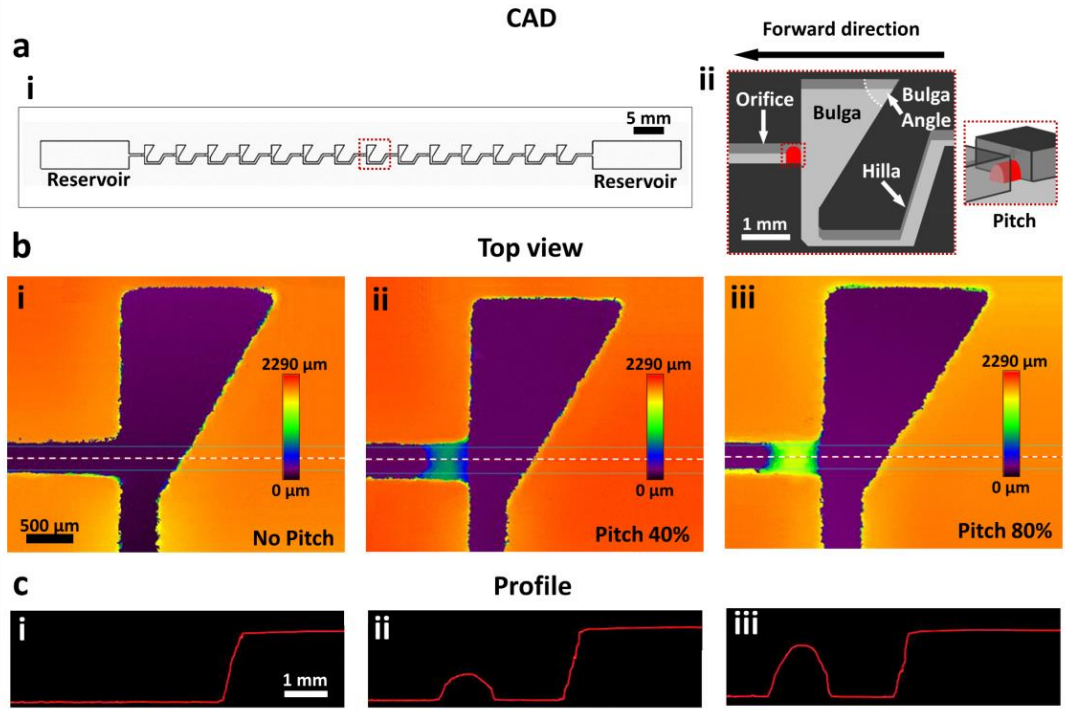

**Fig. 1.** (a) Design of the liquid diodes, showing (i) full channel, with unit cell highlighted in dashed red line, (ii)the main components of a unit cell. (b) Confocal microscope images of unit cells with increasing pitch height, (i) 0%, (ii) 40% and (iii) 80% and (c) corresponding profile images of the pitches from the unit cells shown in (b).

Fig. 1b shows laser confocal micrographs of unit cells with (i) 0%, (ii) 40% and (iii) 80% pitch heights. The corresponding profiles of the pitches are shown in Fig. 1c. One advantage of this design is that the geometry leading to pinning is now only a small section of the channel. This allows us to use the angular expansion of the bulga, shown in Fig. 1a, ii, to adjust velocity with a minimal effect on diode behavior. Both the hilla and the orifice are fixed in length. The length of the hilla is chosen to allow room for varying the bulga angle. The length of the orifice is chosen to ensure that the pinning behavior at the pitch is not influenced by the sloped wall of the hilla.

Diode geometry was identical in both experiments and simulations, with the simulations allowing for finer control over the pitch height and contact angle, and experiments allowing for tests involving many unit cells. In each experiment, the same amount of liquid is placed in the reservoirs, and the flow is optically recorded. The reservoirs enable precise control of the initial conditions in terms of liquid volume and initial pressure within the liquid. Image analysis is used to quantify the flow velocity and diodicity within each unit cell and in the overall channel (see Materials and Methods).

**2.2 Diodic behavior**

We adjust the pitch height and the contact angle (CA) to create an experimental and simulation phase diagram of the fluid flow in the liquid diode (Fig. 2a and 2b). The height of the pitch ranges from 0 to 80% of the total channel depth, in increments of 10%. A full sample features all the nine channels in parallel, with increasing pitch height (Fig. S2, ESI).

The simulations and the experimental results show a close agreement. We first identify the Diodic Regime region (in light blue) in the center of the diagrams. Here, pinning of the liquid in the backward direction is effectively enforced by the combined action of the pitch and the expansion of the channel from the orifice to the bulga. As illustrated in Fig. 2c, the critical pressure required to overcome this pinning is larger than zero for flow in the backward direction, and hence the flow is inhibited without any external driving pressure. The Diodic Regime extends experimentally from CA = 42° to CA = 60° for all pitch heights (Fig. 2d, Video S2, Video S3 and Video S4 in ESI). Below and above these limits, the flow depends on the pitch height.

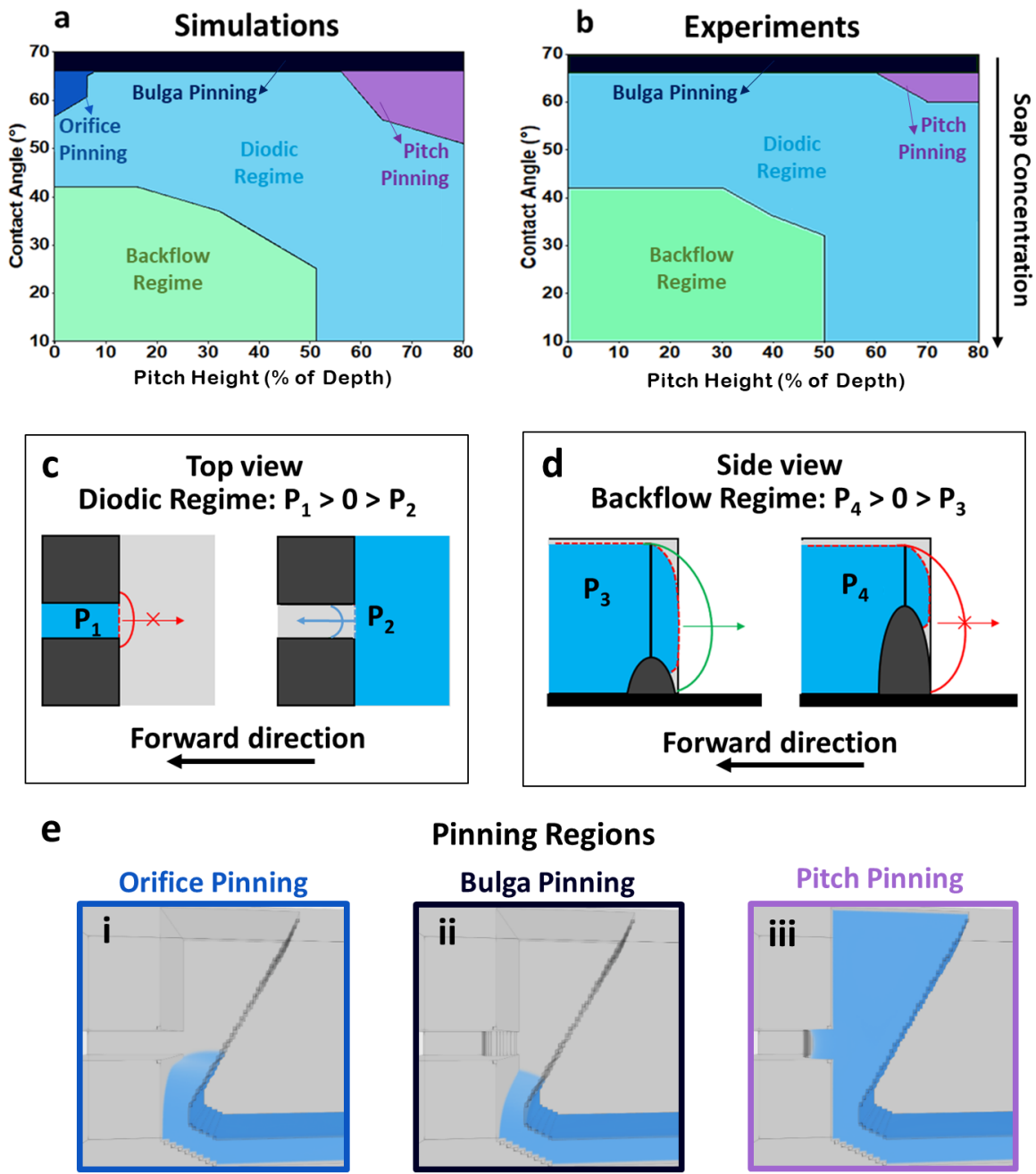

**Fig. 2.** (a) Flow phase diagram for different pitch heights and contact angles, obtained from lattice Boltzmann simulations. Five regimes are found: three upper pinning regions (blue, black and purple) that result in no flow, the diodic region (light blue) and the backward flow region (light green), where diodicity is lost and flow occurs in both directions. (b) Flow phase diagram obtained experimentally. (c) Schematic representation with top view of the diodic principle. The sudden opening in the backward direction creates a pressure barrier that halts the flow. (d) Schematic representation with side view of the diodes' functioning. A higher pitch creates a larger pressure

barrier in the backward direction. (e) Different pinning configurations obtained from the simulations, with (i) no pitch, CA = 60° (Orifice pinning), (ii) 50% pitch height, CA = 70° (Bulga pinning), (iii) 80% pitch height, CA = 60° (Pitch pinning). The colors of the figures' boarders represent different pinning regions according to the phase diagrams.

By reducing the contact angle we enter the Backward Regime, in light green on the bottom left of the Diodic Regime (Fig. 2a and 2b). Here, we observe flow in both directions, as the pressure barrier in the backward direction is lower than the capillary pressure due to the increased wetting of the liquid. The upper boundary of this region starts at CA = 42° and decreases with pitch height, until a pitch height of 50%. As illustrated in Fig. 2d, higher pitches decrease the channel's depth, leading to a more significant widening of the channel's cross-section when entering the bulga in the backward direction, and hence a larger pressure barrier.[29] For pitches higher than 50%, the pressure barrier is high enough to counterbalance even the lowest contact angles (Video S5, ESI).

For contact angles higher than 60°, we still observe experimentally diodic behavior until CA = 65°, for pitch heights 0 – 60%, with the upper limit decreasing for pitch heights of 70% and 80%. Above this, we observe no flow in either direction, for any pitch height. The simulations highlight three different regimes, each characterized by a different pinning configuration, shown in Fig. 2e: Orifice Pinning (in blue) with 0 – 8% pitch height and CA = 55° - 65°, Bulga Pinning (in black) for all pitch heights and contact angles above 70°, and Pitch Pinning (in purple), with pitches between 55% - 80% and CA = 55° - 65°. In the Orifice Pinning region, the pitch is not high enough to act as a surface for the liquid to flow over when filling the bulga. Consequently, the liquid is pinned between the entrance of the orifice and the slanted side of the bulga, forming a convex front (Fig. 2e, i). In the Bulga Pinning region, the liquid gets pinned just after entering the bulga (Fig. 2e, ii), as pressure from the liquid due to capillary forces is not high enough to overcome the slight expansion. Lastly, in the Pitch Pinning region, the liquid fills the bulga and is then halted at the pitch, unable to overcome it (Fig. 2e, iii).

In the experimental phase diagram, the Orifice Pinning region does not appear, and the Pitch Pinning region is smaller than in the simulations. Differences in roughness between simulations and experiments can account for some differences in the phase diagrams.[30] The geometry in the simulations is an approximation to a 3D grid of lattice points and therefore the sloped edge of both the bulga and the pitch are approximated as a staircase rather than a slope. On such geometry, the liquid front requires more energy to advance[31] in comparison to the 3D-printed samples. In the experiments, there are statistical effects related to defects in the samples and degree of cleanness of the 3D-printed surface. Typically, the measured surface roughness of the experimental samples was 32 μm, but there may be variations between the diode walls, floor and pitch. All these may in turn cause variations in the contact angle within the channels and affect the spreading dynamics.[32]

Simulations show as well that varying the bulga angle has little to no impact on diodicity (Fig. S3, ESI). Overall, the pitch alone is what determines the interface between the Diodic and Backflow Regime and the bulga angle can be used to control the speed, as will be discussed in the following sections.

**2.3 Flow patterns**

**2.3.1 High Contact Angles**

In order to understand the flow dynamics, we now present a detailed analysis of the menisci configurations within a single unit cell, for different pitch heights and contact angles. Fig. 3 shows representative flow micrographs from experiments and from simulations in the forward direction, under the diodic regime.

We distinguish between different flow regimes according to the pitch height. For instance, for no pitch the liquid pins briefly, forming a convex liquid front between the orifice's entrance and the slanted side of the bulga (Fig. 3a, t = 0 s and t = 1.1 s). Shortly after, flow is resumed in the forward direction, while the bulga is still being filled (Fig. 3a, t = 1.76 s). For mid-range pitches (20% - 50%), the liquid pins at the pitch, forming a convex liquid from inside the bulga, as already seen in Fig. 2c, ii. The liquid front remains pinned until the bulga is completely filled with enough volume to overcome the pitch. For higher pitches (60% - 80%), we see similar dynamics, but differences in the flow timescales. In each case, the bulga must eventually be filled for the liquid front to advance. Namely, the volumetric flow rate of the channels is determined by the volume of the bulga. This principle allows velocity to be controlled by adjusting the bulga angle. Additionally, the expansion angle of the bulga determines the capillary pressure required for the liquid to advance into the top corner.[33] There are, therefore, two mechanisms by which the bulga can control the flow speed.

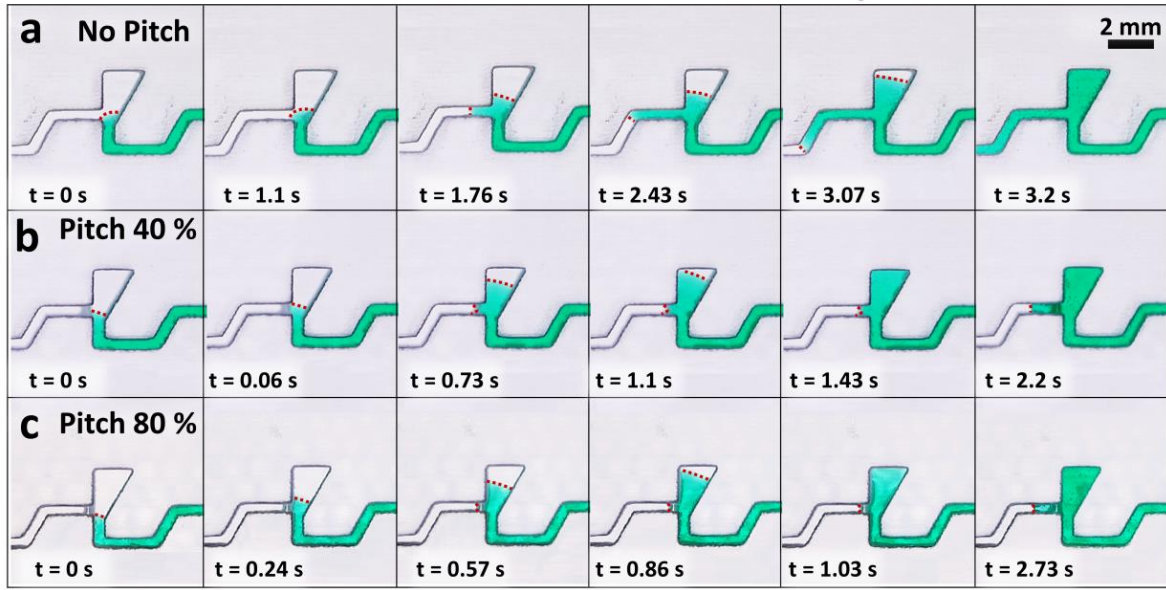
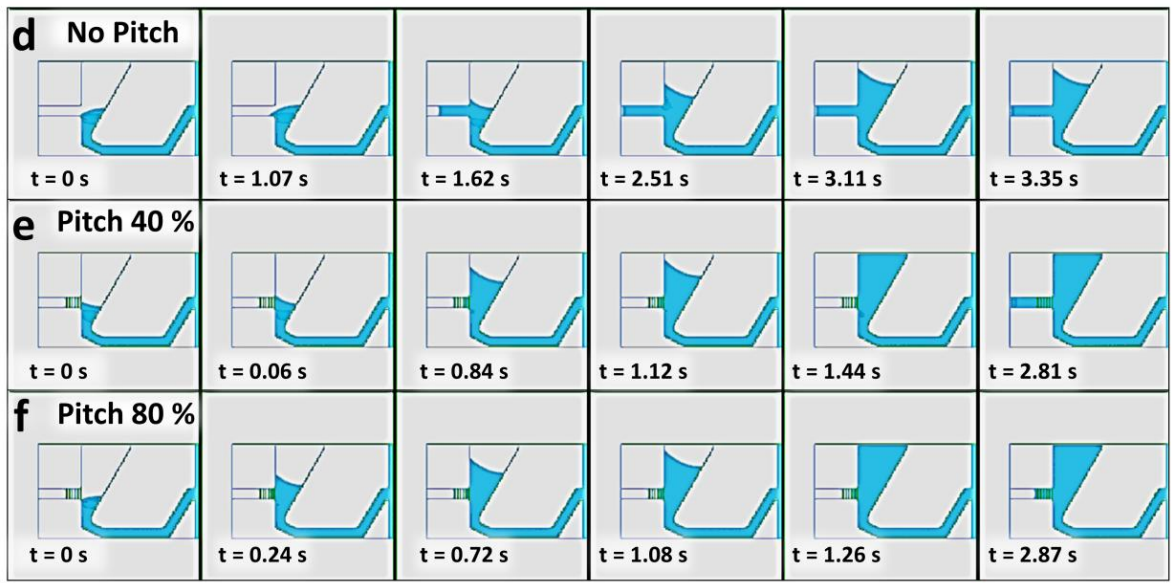

**Fig. 3.** Images of flow patterns inside individual unit cells, obtained from (a, b, c) experiments and (d, e, f) simulations for CA = 50° ± 3° and pitch heights of 0%, 40% and 80%. With no pitch, the liquid flows into the orifice before beginning to fill the bulga. For 40% and 80% pitch heights, liquid fills the bulga up to the height of the pitch and then proceeds to enter the orifice.

**2.3.2 Low Contact Angles**

For lower contact angles, geometric features of the bulga become more significant for flow. The liquid front in the corners of the bulga advances more quickly than the bulk, as predicted by the Concus-Finn condition.[34,35] This condition describes the increased wetting observed in interior corners for low liquid-solid surface tensions. It occurs with contact angles, *CA*, and corner half-angles, $\alpha$, following:[36]

$$CA < \pi/2 - \alpha. \tag{1}$$

The liquid diode design in this study has several sharp corners. The floor and the walls of the diode form a 90° angle, meaning that contact angles below 45° will show Concus-Finn behavior. Additionally, the bulga has a maximum angle of 80° in our tests, meaning that contact angles below CA = 50° will have preferential wetting in this corner, accelerating flow. This is demonstrated in Fig. 4, where we observe significant wetting in the corners for CA = 25°, a bulga angle of 60° and a pitch height of 40%, both in experimental results (Fig. 4a) and in simulations (Fig. 4b). The bulga angle is defined as shown in Fig. 1a, ii. Consequently, the time taken to fill the bulga completely is reduced compared to higher contact angles, as the liquid will be pulled up the acute corner of the bulga by capillary forces. This can be seen when comparing Fig.s 3b and 3e to Fig.s 4a, i and 4a, ii, respectively. The Concus-Finn effect applies to a wider range of contact angles for smaller bulga angles. For instance, for a bulga angle of 40°, we will observe this behavior for the entire diodic range of contact angles (10° - 65°).

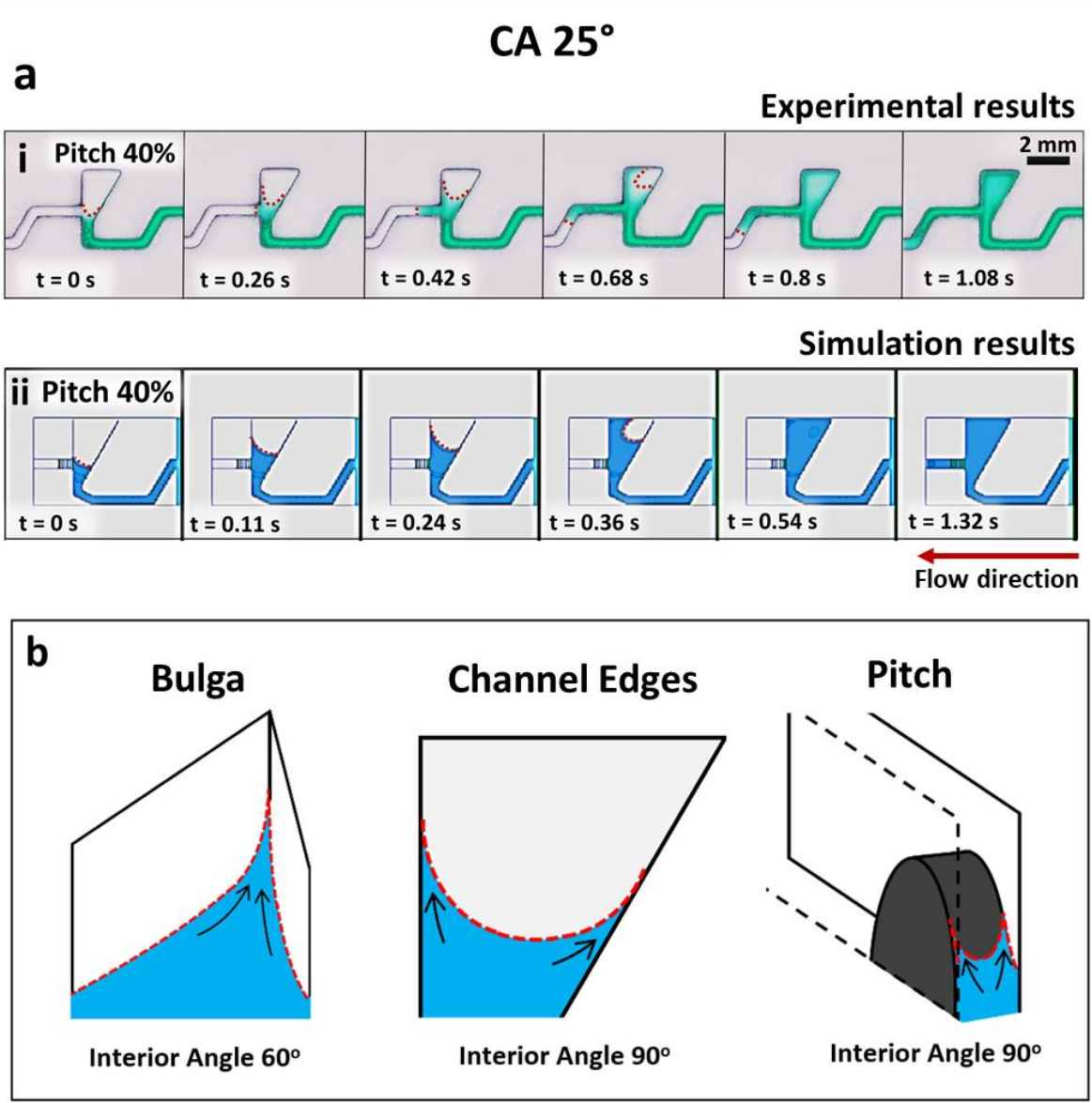

**Fig. 4.** (a) Concus-Finn behavior observed in (i) experiments and (ii) simulations, on surfaces with CA = 25°. The liquid front advances significantly faster in the corners, with greater curvature in comparison to the flow in Fig. 3, with CA = 50°. (b) Schematic representation of the wetting of the bulga's interior corner (left), the channel edges within the bulga (center) and the pitch (right) with CA = 25°, showing Concus-Finn behavior.

In the experiments, flow resumes into the orifice before the bulga has been filled, in contrast to the simulations. In fact, due to its increased wetting, the liquid can climb up the pitch without needing the bulga to be filled. In the simulations, however, the staircase-like 3D grid approximation makes it harder for the liquid to climb the pitch and, consequently, the bulga is filled before flow resumes.

## 2.4 Flow Dynamics

We now determine the governing laws of imbibition in the channels within the Diodic Regime (light blue region in Fig. 2a-b). Our findings are summarized in Fig. 5. Fig. 5a, i-ii illustrates position and velocity plots for flows with CA = 43° ± 3° and pitch height of 0%, 10% and 80%. On the right, side-view CAD sketches of a unit cell elucidate the variation of the pitch height (Fig. 5a, iii-v). Fig. 5b shows the same plots as Fig. 5a, but for three different bulga angles, 80°, 60° and 40°, with CA = 43° ± 3° and pitch height of 50%. On the right, top view CAD sketches of a unit cell elucidate the variation of the bulga angle (Fig. 5b, iii-v).

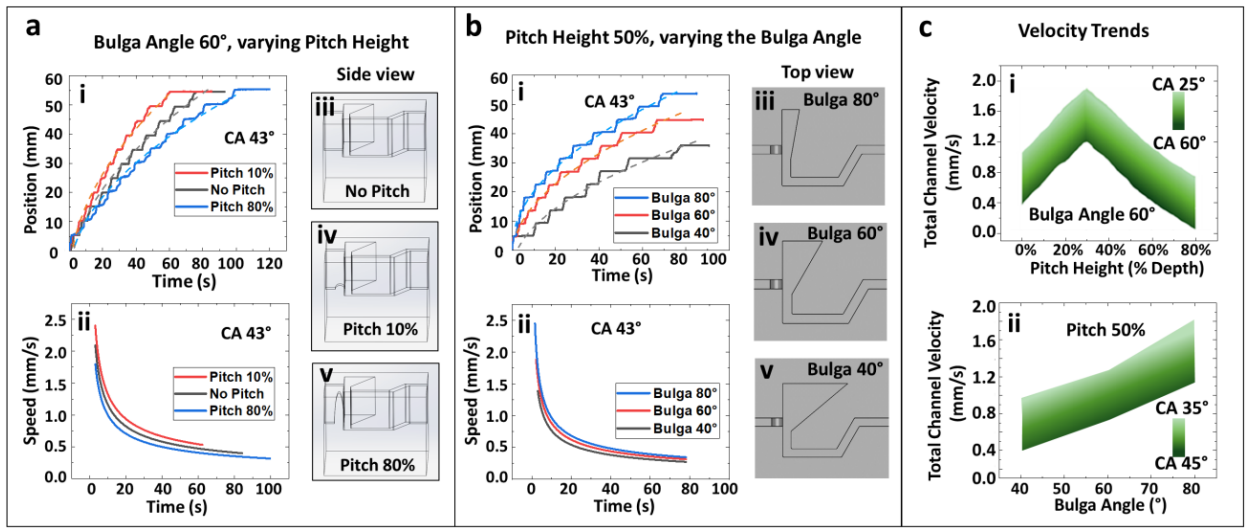

**Fig. 5.** Flow dynamics in the liquid diodes. (a) Liquid front (i) position and (ii) speed in the channel over time for pitch heights of 0%, 10% and 80%, bulga angle = 60° and CA = 43° ± 3°; (iii-v) sketches of unit cells with 0%, 10% and 80% pitch heights, respectively. (b) Liquid front (i) position and (ii) speed in the channel over time for bulga angle = 80°, 60° and 40°, pitch height =

50% and CA = 43° ± 3°; (iii-v) sketches of unit cells with 80°, 60° and 40° bulga angle, respectively. (c) Total channel velocity trends as a function of (i) pitch height and contact angle, for bulga angle = 60° and (ii) bulga angle and contact angle, for pitch height = 50%.

When examining the advancement of the liquid front into the channels, we observe a time-dependent step-like behavior (Fig. 5a, i and 5b, i). Each complete step corresponds to the filling of a single unit cell. As the position of the liquid front is measured only in the direction parallel to the orifice, the horizontal lines of the steps represent the liquid being temporarily pinned while the bulga is being filled, or before the liquid has enough driving force to proceed into the orifice in the case of 0% pitch height, as explained in section 2.2. The size of the steps increases along the channel, meaning that filling each following unit cell takes more time than the previous one.

The unidirectional transport distance in the diodes follows a typical Lucas-Washburn law. This is due to the increasing friction (viscous dissipation) of the wetting liquid in the channel with increasing volume of the liquid transported.[37,38] In Fig. 5 a-b, i Lucas-Washburn fits of our data are represented by dashed lines, following:

$$x(t) = at^{\frac{1}{2}} + c. \qquad (2)$$

*x(t)* is the position of the advancing liquid front, *a* is a parameter that characterizes the flow speed and depends on the characteristics of the system, namely the pitch height, contact angle, viscosity of the fluid and surface roughness. The constant *c* accounts for initial dissipation by the reservoirs. For details of the best fit parameters, see Fig. S5 in the ESI. The approximation is less accurate for the first unit cell. We ascribe this divergence from the Lucas-Washburn law to increased pressure from the initial filling of the reservoirs.

Once the fit to Eq. (2) is done, we can take a time derivative to analyze the fluid flow velocity, as shown in Fig. 5a-b, ii. As we use this procedure, the plots do not exhibit any step-like behavior, but rather show an average of each step. For each case, the velocity decreases over time as $t^{-1/2}$. Comparing the velocity for different diode designs, we observed three main trends (Fig. 5c). The first one relates to the pitch height. As already evident from Fig. 5a, i-ii, channels with 10% pitch height are faster than channels with no pitch, as well as those with 80% pitch height. Indeed, velocity increases at first as pitch height increases, reaching its maximum at a pitch height of 30% before dropping to its minimum at a pitch height of 80%. This is shown in Fig. 5c, i, which provides a color-code map of the liquid front average velocity over the entire channel, as a function of pitch height and contact angle. This velocity trend is a direct result of the different flow patterns we see for different pitch heights (Fig. 3). The pitch appears to reduce the resistance on the liquid when entering the orifice but also increases the volume required to fill the bulga, causing the flow to momentarily halt in each unit cell. The side of the pitch provides the liquid with an additional surface for capillary forces to act over, explaining why velocity increases initially for increasing pitch height. However, for pitches above 30% in height, this additional thrust is not enough to counterbalance the increased time needed to fill the bulga. Consequently, after this threshold, velocity decreases with pitch height.

The second velocity pattern relates to the bulga angle. We observed an increase in the velocity with increasing bulga angle, meaning smaller bulga's volume. This is illustrated in Fig. 5c, ii, which features a color-code map of the liquid front average velocity over the entire channel, as a function of bulga angle and contact angle, for a pitch height of 50%. Indeed, with less volume to fill, the steps observed in the liquid front position graph (Fig. 5b, i) become smaller and the slopes become steeper. Thus, times are reduced, and we observe an increase in the liquid front's initial velocity of 30% and 35% at each 20° increase of the bulga angle (Fig. 5b, ii). This trend is independent of the pitch height. Thus, by adjusting the bulga angle and the pitch height we can effectively tune the velocity of the channels to optimal performance.

Lastly, as expected, increasing the contact angle also causes a reduction in both the instantaneous velocity and the overall channel speed. This is seen both in Fig. 5c, i and 5c, ii and is a direct consequence of poorer wetting between the liquid and the channels' walls with higher contact angles, resulting in less driving force for the flow.[39] A plot showing the dependence of the flow velocity on the contact angle, for three representative pitch heights with 60° bulga angle is provided in the ESI (Fig. S6).

**2.5 Condensation from Humid Air**

We employed the liquid diodes for collecting and directing condensed water from humid air. Fig. 6a, i shows the experimental setup, consisting of a humidifier, a fog chamber and the diodes. The diodes are inserted into the chamber through a slot, while the reservoirs remain outside. It is possible to tilt the channels in order to test flow uphill, as shown Fig. 6a, ii. The fog from the humidifier condenses on the reservoirs' surface and flows spontaneously through the channel. Fig. 6b shows an experiment using a channel with pitch height of 40% and tilting angle of 10°. The condensed liquid successfully climbs up the channel, eventually reaching the other end. Fig. 6c and 6d show condensation experiments with the same pitch height as Fig. 6b but without tilting, in the forward and backward directions, respectively. The latter was achieved by rotating the samples by 180° and inserting it into the fog chamber, with the reservoirs to the left of the samples (Fig. 1a, ii). For pitches of up to 50% in height, the liquid enters the orifice before filling up the bulga for the first 3-4 unit cells (Fig. 6c, t = 30 min), contrary to what we observed in previous experiments. This is attributed to the dropwise condensation, where the vapor condenses first in small separate droplets, that then grow bigger through coalescence.[40] Hence, in the initial stages of the experiments, the reservoirs are not filled homogeneously. Once droplets join together to fill the reservoirs, the expected flow patterns are observed again. Both the tilted and horizontal surfaces successfully transport approximately 100 µL of condensed liquid through the channels. However, in the absence of gravitational forces, the horizontal sample transports the liquid faster, as expected. Namely, the tilted channel is filled completely within 80 min, while the horizontal chamber is filled after 60 min. In the backward direction, no flow was observed, both without and with tilting of the sample.

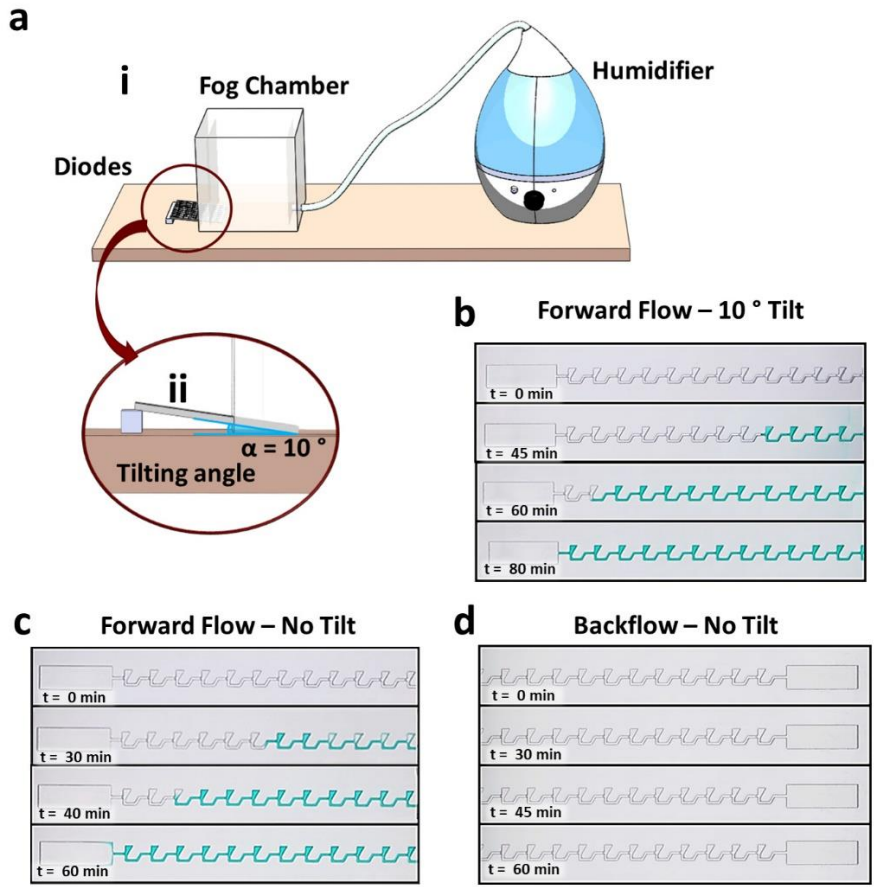

**Fig. 6:** (a) Condensation experiments, consisting of (i) a humidifier, a fog chamber and the diodes, inserted into the chamber through a slot, up until the reservoirs. The channels can be tilted (ii). (b) Optical images of condensed aqueous dyed solution with CA = 50°. The channels have pitch with 40% height and tilting angle of 10°. (c-d) Condensation in a channel with pitch height of 40%, without tilting in the (c) forward and (d) backward directions.

## 3. Conclusions

In this work, we elucidate the role of design parameters and wettability in directional liquid transport in open-channel liquid diodes. Consequently, we are able to control the pressure barrier for transport and speed of the fluid flow in the channels. We focus on three main parameters. First, an ellipse shaped pitch, which can be adjusted to control the cross-sectional area of the sudden widening channel, used to enforce diodic behavior. Second, a reservoir within each unit cell of the diode, named the bulga, that is used to tune the liquid velocity in the channel. Third, the contact angle of the liquid-solid interface.

The pitch can be introduced to other liquid diodes to strengthen diodic behavior and widen their working window. Velocity increases with pitch height, up to a maximal value of roughly 2 mm/s, for pitch height of 30% of the channel depth. Above this pitch height, velocity decreases. Overall, velocity increases up to roughly three times in diodes with pitch in comparison with diodes with no pitch at all. In addition, the shape of a reservoir in each unit cell, the bulga, can be tuned to increase the flow average velocity up to 50% with no impact on diodic behavior. In terms of contact angle, there is an interplay between the driving force for the liquid to spread for lower contact angles and the collapse of the diodicity. Simulations show there are several pinning mechanisms of the liquid within the diodes, depending on the design and wettability of the channels.

3D-printing, which remains an underexploited manufacturing method for liquid diodes, offers an accessible and economic way not only to highlight the role of geometrical parameters in directional fluid transport, but also to design ad-hoc devices for accumulation and directional spontaneous transport of fluids. Our channels successfully demonstrate their ability to harvest fog, while transporting the condensed liquid even against gravity. In the future, it will be interesting to study more complex designs. For instance, the flow patterns in two-dimensional networks (Fig. S7, ESI).[41] Another fruitful direction is to consider more complex fluids, in particular as many liquids of interest for liquid diode applications can have non-Newtonian rheological behaviors.[42]

## 4. Materials and Methods

### 4.1 Liquid diode design

Liquid diodes were designed using SolidWorks. The size ratios used in the design are the same ones used in the simulations.

### 4.2 Sample Preparation

We fabricated the samples by 3D-printing, using the ProJet® MJP 2500 Series by 3D Systems (Rock Hill, South Carolina, USA), a multi-jet 3D printer. The material used for the printing is the VisiJet® M2R-CL, a transparent plastic. Four different samples were printed, to check repeatability of the results.

After printing, the parts were cleaned to remove the supporting wax material. First, the support wax is dissolved in canola oil at up to 65 °C under stirring or mechanical shaking. The excess oil, with wax residues dissolved in it, is then removed with a Kimwipe and the sample is placed in new clean oil at 65 °C. When the wax is completely removed, the oil residues are washed in soapy water (all-purpose liquid detergent soap, ratio soap to water 0.3:1) at 65 °C under stirring or mechanical shaking. The cleaning is finalized by using gentle brushing in order not to damage small features. Lastly, soap residues are thoroughly rinsed in DI water and the sample is left to dry in open air.

To conduct experiments, we used DI water, to which we added green food coloring (Maimon's, Be'er Sheva, Israel) in order to provide contrast for the subsequent video analysis process. To decrease the native contact angle of the liquid, we added small amounts of soap, the same using during cleaning, from a ratio of 0.05:1 to a ratio of 0.1:1.

### 4.3 Characterization

*4.3.1 Confocal microscopy*

The morphology of the samples was characterized via laser confocal microscopy, using the LEXT™ OLS5100 by Olympus Corporation (Shinjuku City, Tokyo, Japan).

*4.3.2 Contact angle measurements*

Wettability of the different liquids used was assessed via contact angle measurements, using the OCA25 by DataPhysics Instruments GmbH (Filderstadt, Germany).

For each measurement, five drops of liquid (2 μL) from different areas of the surface were analyzed.

### 4.4 Flow Experiments and Image Analysis

To conduct experiments, we placed an 80 μL drop of liquid of known contact angle in each reservoir of each channel. We used Transferpette® S pipettes by BRAND GMBH + CO KG (Wertheim, Germany). This volume ensures filling of 12 unit cells, while neutralizing the effect of pipette-induced pressure. We repeated each experiment three times on different samples and recorded it from above using a camera (Panasonic DC-S1 with Sigma 70 mm f/2.8 DG Macro lens). Average velocities were calculated by dividing the total flow time by the length covered by the liquid.

The videos were analyzed using a MATLAB script. The algorithm separates each frame into three RGB channels and subtracts the green from the red, which helps highlight only the liquid front with respect to the sample background. Each frame is then turned into a binary black and white image and the final image sequence, showing only the advancing liquid front, is obtained. The position of the liquid front at each instant in time is then recorded.

OriginPro Software was used to produce position and velocity graphs. Velocities were obtained from fitted curves by means of first derivatives.

For condensation experiments, we used the setup shown in Fig. 6a, i-ii. We used an Ultrasonic Humidifier, operating at 25 W and 50 Hz. We inserted the diodes into the fog chamber, through the dedicated slot, only until the reservoirs. To conduct experiments in the forward direction, we inserted the reservoirs on the right-hand side of the channels. For the backward direction, we rotated the sample by 180º, thus inserting the reservoirs on the left-hand side. To test the flow uphill, we placed a spacer under the edge of the sample outside the chamber. The humidifier was filled with a liquid of known contact angle and turned on at maximal power. Experiments were recorded from above.

### 4.5 Simulations

*4.5.1 Numerical Method*

Simulations were performed using the lattice Boltzmann method[43] to study the capillary dynamics. The system evolves according to the continuity equation (3), the Navier-Stokes equation (4) and the Cahn-Hilliard equation (5).

$$\frac{\partial \rho}{\partial t} + \nabla \cdot (\rho \vec{u}) = 0, \tag{3}$$

$$\rho \frac{\partial \vec{u}}{\partial t} = -\nabla P + \eta \nabla^2 \vec{u} + \vec{F}, \tag{4}$$

$$\frac{\partial \phi}{\partial t} + \nabla \cdot (\phi \vec{u}) = \nabla \cdot (M \nabla \mu). \tag{5}$$

$\phi$ is an order parameter that distinguishes the two phases, with $\phi = 1$ for the liquid and $\phi = -1$ for the gas. Velocity is represented by $\vec{u}$, and $\rho$ is the density. $\eta$ represents the shear viscosity and $M$ is a mobility parameter which is taken to be constant. The chemical potential $\mu$ and pressure tensor $P$ can be computed following the free energy of the system.[44] For our system, The free energy is given by[44,45]

$$\Psi = \int_V \left[ c_s^2 \rho \ln \rho + \frac{A}{4}(\phi^2 - 1)^2 + \frac{\kappa}{2}(\nabla \phi)^2 \right] dV - \int_A h \phi_s \, dA. \tag{6}$$

The chemical potential and pressure tensor can then be derived as

$$\mu = -A\phi + A\phi^3 - \kappa \Delta \phi, \tag{7}$$

$$\partial_\beta P_{\alpha\beta} = \partial_\alpha (c_s^2 \rho) + \phi \partial_\alpha \mu, \tag{8}$$

where $c_s$ represents the speed of sound, the parameter $A$ is used to adjust the width of the diffuse interface between the two components, and $\kappa$ relates to the surface tension between the liquid and gas. A surface free energy term to account for interactions between the fluids and the solid interface is also included, with the parameter $h$ related to the contact angle $CA$ between the liquid and the solid

$$h = sgn(\pi/2 - CA)\sqrt{2\kappa A}\sqrt{cos(\alpha/3)[1 - cos(\alpha/3)]}, \qquad (9)$$

$$\alpha(CA) = arccos(sin^2(CA)). \qquad (10)$$

The $sgn$ function returns the sign of the expression in brackets and $\alpha$ represents the width of the interface between the liquid and gas components.

For the capillary flows considered here, it is important that we capture suitable contact line dynamics. There are two main models for contact line dynamics, based on hydrodynamics theory and molecular kinetic theory.[46,47] Our modelling approach here aligns with the hydrodynamics theory. For example, our previous works have shown that this approach can correctly reproduce the Cox-Voinov's law for the dynamic contact angle,[48] as well as stick-slip motion when there are chemical and topographical heterogeneities on the surface.[49]

*4.5.2 Diode Geometry*

The design used in the simulation is the closest approximation of this that can be represented on a discrete 3D lattice with a finite resolution. Boundary conditions were enforced using a no-slip condition, ensuring that velocity on the boundary parallel to the wall was zero.

The contact angles and pitch heights can be varied easily in the simulations. The depth of the diode was 50 lattice units, so pitch height could be varied in intervals of 2% in each simulation. Contact angles could be adjusted freely. This allowed for a detailed insight into how the contact angle and pitch height affected diode behavior.

*4.5.3 Flow Measurements*

The liquid interface was tracked, and flow in either direction was deemed to occur when this passed the furthest point of the pitch from the reservoir. Equally, this is the point at which the liquid has cleared the pitch and continued to flow in either direction. In contrast to this, we determined cases with no flow to occur when the maximum velocity change between successive iterations fell below 0.01 mm/s. For the phase diagram tests, only one unit cell was used, with a reservoir initialized on either side of the diode to investigate flow in either direction.

*4.5.4 Conversion to Real Time*

Simulation iterations can be related to real time by matching the length scale, mass and surface tensions to the values used in experiments. We can then use the relations between these quantities to calculate the time step in seconds.

**Supplementary Material**

Video S1: Flow experiment over long distance. Videos S2-S5: Videos showing experimental and simulated flow through the diode in two directions with a variety of contact angles and pitch heights. Video S6: Condensation experiment video. Fig. S1-S3: Supplementary Figures showing relations between pitch shape and diode behavior, the impact of bulga angle and orifice height on diode behavior and flow time and the full 3D-printed sample. Fig. S5: Parameters of fitted curves

following the Lucas-Washburn law from flow experiments. Fig S6: Dependence of flow velocity on contact angle. Fig. S7: Simulation results for two diodes connected perpendicular to each other.

**Author Contributions**

C.S. developed the experimental setup and carried out the experiments. M.R. developed and carried out the simulations. H.K and B.-E.P conceived the research project and provided supervision. All authors discussed the results and contributed to the writing of the manuscript. All authors have given approval to the final version of the manuscript.

ACKNOWLEDGMENTS

This research was supported by the Israel Science Foundation (Grant No. 1323/19), by the Carl Gans Foundation, and by the UK Research and Innovation (UKRI) Engineering and Physical Sciences Research Council (EPSRC) (Grant No. EP/V034154/1). We thank Dr. Chenyu Jin and Dr. David Feldmann for assisting in writing the MATLAB script used for image analysis, Uriel Naor and Raz Samira for assistance with confocal microscopy and Leeor Mordorch for assistance with SolidWorks.